\documentclass[aps,prev,twocolumn,preprintnumbers,floatfix,nofootinbib]{revtex4-1}

\usepackage[T1]{fontenc} 
\usepackage[utf8]{inputenc}
\usepackage{physics}
\usepackage{subfigure}

\usepackage{amssymb,graphicx}
\usepackage{mathrsfs}
\usepackage{verbatim}
\usepackage{color}
\usepackage[dvipsnames]{xcolor}
\usepackage{multirow}
\usepackage{amsmath}

\usepackage{slashed} 
\usepackage{float}
\usepackage[normalem]{ulem}
\usepackage{sidecap}
\usepackage{hyperref}
\usepackage{cancel}
\usepackage{enumerate}
\usepackage{array}
\hypersetup{pdfstartview=FitV,colorlinks=true,linkcolor=blue,citecolor=red,filecolor=black,urlcolor=blue}

\newcommand{\aend}{a_{\rm end}}
\newcommand{\arh}{a_{\rm RH}}
\newcommand{\rhorh}{\rho_{\rm RH}}
\newcommand{\rhoend}{\rho_{\rm end}}
\newcommand{\beq}{\begin{equation}}
\newcommand{\eeq}{\end{equation}}
\newcommand{\bea}{\begin{eqnarray}}
\newcommand{\eea}{\end{eqnarray}}

\def\trh{T_{\rm RH}}

\usepackage[parfill]{parskip}
\begin{document}\sloppy

 \preprint{UMN--TH--4311/24, FTPI--MINN--24/03}

\vspace*{1mm}

\title{Minimal Production of Prompt Gravitational Waves during Reheating}
\author{Gongjun Choi}
\email{choi0988@umn.edu}
\author{Wenqi Ke}
\email{wke@umn.edu}
\author{Keith A. Olive}
\email{olive@umn.edu}
\vspace{0.5cm}
\affiliation{William I. Fine Theoretical Physics Institute, School of
 Physics and Astronomy, University of Minnesota, Minneapolis, MN 55455,
 USA}

\date{\today}

\begin{abstract} 
The inflationary reheating phase begins when
accelerated expansion ends. As all Standard Model particles are coupled to gravity, gravitational interactions will lead to particle production. This includes the thermal bath, dark matter and gravitational radiation. 
Here, we compute the spectrum of gravitational waves from the inflaton condensate during the initial phase of reheating. As particular examples of inflation, we consider the Starobinsky model and T-models, all of which are in good phenomenological agreement with CMB anisotropy measurements. The T-models are distinguished by the shape of the potential about its minimum and can be approximated by $V \sim \phi^k$, where $\phi$ is the inflaton.   Interestingly, 
the shape of the gravitational wave spectrum (when observed) can be used to distinguish among the models considered. As we show, the Starobinsky model and T-models with $k=2$, provide very different 
spectra when compared to models with $k=4$ or $k>4$. Observation of multiple harmonics in the spectrum can be interpreted as a direct measurement of the inflaton mass. 
Furthermore, the cutoff in frequency can be used to determine the reheating temperature. 

\end{abstract}

\maketitle

\setcounter{equation}{0}

\section{Introduction}

The basic inflationary paradigm \cite{reviews} makes several testable predictions. For example, it is expected that the universe is flat \cite{Guth}; probably to very high precision. That is, the total energy density relative to the critical energy density, $\rho_c$, is very close to unity, $\Omega = 1$. Generally inflationary models predict a spectrum of density fluctuations close to but slightly redder than the Harrison-Zeldovich spectrum\cite{Harrison:1969fb,Zeldovich:1972zz}, i.e., a scalar spectral index, $n_s \lesssim 1$ \cite{Mukhanov:1981xt}. Models also predict a non-vanishing ratio of tensor to scalar perturbations, though the exact value is more model dependent. The first two of these three predictions have been borne out by experiment, primarily observations of cosmic microwave background (CMB) anisotropies \cite{wmap,Planck}. Limits on the tensor-to-scalar ratio  \cite{BICEP2021,Tristram:2021tvh} have been able to exclude some simple models of inflation. 

Additional signatures and experimental verification of inflation and inflationary models are clearly needed. One possible signature is the production of gravitational waves during reheating \cite{Grishchuk:1993te,Turner:1996ck,Dufaux:2007pt,Easther:2007vj,Assadullahi:2009nf,Brandenberger:2011eq,Cook:2011hg,Ema:2015dka,Ema:2016hlw,Kuroyanagi:2017kfx,Nakayama:2018ptw,Huang:2019lgd,Ema:2020ggo,Mishra:2021wkm,Haque:2021dha,Barman:2023ymn,Chakraborty:2023ocr,Barman:2023rpg,CrispimRomao:2023fij,Kanemura:2023pnv,Bernal:2023wus,Tokareva:2023mrt}. Recently, several works have considered the Bremsstrahlung production of gravitational waves from inflaton decay \cite{Nakayama:2018ptw,Huang:2019lgd,Barman:2023ymn,Chakraborty:2023ocr,Kanemura:2023pnv,Bernal:2023wus,Tokareva:2023mrt}. This is dominated by processes at the end of reheating. But it is also possible to produce gravitational waves from scattering within the inflaton condensate \cite{Ema:2015dka,Ema:2016hlw,Ema:2020ggo}. Since these processes depend on the (square of) the inflaton density, the production of these waves are dominated by scatterings at the onset of reheating as the oscillatory phase begins. 

Recognizing that reheating is not an instantaneous event \cite{Giudice:2000ex,Bernal:2020gzm,GMOP}, there has been considerable interest in particle production during the reheating period \cite{GMOP,Bernal:2020gzm,egnop,KMO,GKMO1,GKMO2,Bernal,Bernal:2019mhf,Barman:2020plp,Chen:2017kvz,moz,Becker:2023tvd,frag,gkkmov}. 
Many of these processes are related to the so-called freeze-in mechanism \cite{fimp,Bernal:2017kxu} for which the gravitino is a prime example \cite{nos,ehnos,kl}.
There are also processes which rely only on gravitational interactions \cite{Ema:2015dka,Ema:2016hlw,Garny:2015sjg,ema,Tang:2017hvq,Bernal:2018qlk,Chianese:2020yjo,Kolb:2020fwh,Redi:2020ffc,Ling:2021zlj,MO,Barman:2021ugy,Ahmed:2021fvt,Bernal:2021kaj,Haque:2021mab,cmov,Haque:2022kez,Aoki:2022dzd,cmosv,cmo,Ahmed:2022tfm,moz,Basso:2022tpd,Barman:2022qgt,Haque:2023yra,Kaneta:2022gug,Kolb:2023dzp,Kaneta:2023kfv,Garcia:2023qab,Zhang:2023xcd,RiajulHaque:2023cqe,kkmov,gkkmov}. Here we will concentrate on the direct production of gravitational waves 
from scattering within the inflaton condensate. As these processes are unavoidable (they rely only on the coupling of the inflaton to gravity), they are a minimal source of gravitational waves present in inflationary theories. In particular, they do not rely directly on the couplings of the inflaton to matter leading to inflaton decay, though the final frequency spectrum 
of gravitational waves will depend on the reheating temperature as it will affect the degree to which the frequencies redshift from the moment of their creation to the present time. 

As concrete examples, we will consider both the Starobinsky model of inflation \cite{Staro}
with a scalar potential given by
\beq
V(\phi)  \; = \;  \frac34 m_\phi^2 M_P^2 \left(1 - e^{-\sqrt{\frac{2}{3}} \frac{\phi}{M_P}} \right)^2 \, ,
\label{staropot}
\eeq
where $\phi$ is the inflaton, $M_P\simeq 2.4\times 10^{18}\text{ GeV}$ is the reduced Planck mass and the related $\alpha$-attractor T-models~\cite{Kallosh:2013hoa} with scalar potential
\beq
\label{eq:Vattractor}
V(\phi) \;=\; \lambda M_P^4 \left[ \sqrt{6} \tanh\left(\frac{\phi}{\sqrt{6} M_P }\right)\right]^k\,.
\eeq
Both examples are
phenomenologically viable \cite{egnov}.
When the T-model potential is expanded about its minimum, the potential can be approximated by 
\begin{equation}
    V(\phi)=\lambda M_P^4 \left( \frac{\phi}{M_P}\right)^k \,, \quad \phi\ll M_P \,.
    \label{kexp}
\end{equation} 
For $k=2$, the T-models as well as the Starobinsky model, can be approximated by a quadratic potential providing a source of harmonic oscillations of the inflaton condensate when the period of inflationary expansion ends. For $k\ne 2$, the T-models give rise to anharmonic oscillations.  

In analogy with the gravitational production of matter produced by inflaton scattering in \cite{MO,cmov,kkmov,gkkmov}, here we consider the pair production of gravitons from inflaton scattering. In Section \ref{PR},
we provide our computation of the amplitudes for the processes considered with some details reserved for the Appendix~\ref{app:feynrules}. 
These are employed in Section \ref{spect} and we solve the Boltzmann equations for the simple case of the Starobinsky potential as well as the T-model potential with $k=2$  for the production of the gravitational wave spectrum. In Section \ref{kne2}, we generalize to the T-models with $k > 2$. Our conclusions are given in Section \ref{sec:sum}.

\section{Production Rates}
\label{PR}
We consider the minimal production of gravitational waves from inflaton annihilation, independent of the matter couplings. We start from the Einstein-Hilbert action and the inflaton ($\phi$) action minimally coupled to gravity:
\begin{equation}
    S=\int d^4 x \sqrt{-g} \left(2\kappa^{-2}R+\frac{1}{2}g^{\mu\nu } \nabla_\mu\phi\nabla _\nu  \phi-V(\phi)\right) \,,
\end{equation}
where $\kappa\equiv \sqrt{32\pi G}=2/M_P$. In the weak field limit, the metric $g_{\mu\nu}$ can be expanded in powers of $\kappa$ as:
\begin{equation}
   \begin{aligned}
       &g_{\mu\nu}=\eta_{\mu\nu}+\kappa h_{\mu\nu}+\cdots\\&g^{\mu\nu}=\eta^{\mu\nu}-\kappa h^{\mu\nu}+\kappa^2 h^{\mu\lambda}h_{\lambda}{}^\nu+\cdots\,,
   \end{aligned} \end{equation}
and $h_{\mu\nu}$ is identified with the canonically normalized graviton. The first order of the expansion yields the scalar coupling to a graviton (see e.g., \cite{hol}):
\begin{equation}
    \mathcal{L}\supset -\frac{\kappa}{2}  h_{\mu\nu} T_\phi^{\mu\nu}\,,
\end{equation}
where $T_\phi^{\mu\nu}=\partial^\mu \phi\partial^\nu \phi-g^{\mu\nu}\left[\frac{1}{2}\partial^\alpha \phi\partial_\alpha\phi -V(\phi)\right]$ is the scalar energy-momentum tensor.
The quartic vertex $hh\phi\phi$ arises at second order with a coupling constant $\kappa^2$. Besides, in this expansion, the Ricci scalar   gives rise to graviton self-interactions, where the trilinear vertex comes with a coefficient $\kappa$. With this minimal content, the dominant source of graviton production is  through inflaton annihilation, by inflaton exchange, graviton exchange, or the contact term. 

\begin{figure*}[htp]
  \centering
  \hspace*{-5mm}
  \subfigure[]{\includegraphics[scale=1.2]{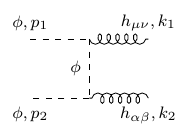}}\quad
  \subfigure[]{\includegraphics[scale=1.2]{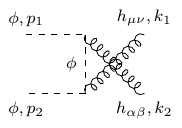}}\quad
  \subfigure[]{\includegraphics[scale=1.2]{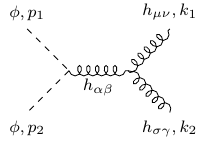}}\quad
  \subfigure[]{\includegraphics[scale=1.2]{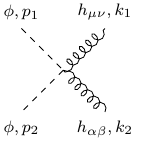}}
  \caption{Feynman diagrams for graviton pair production from inflaton annihilation. The momenta $p_1, p_2$ are incoming and $k_1, k_2$ are outgoing.}
  \vspace*{-1.5mm}
\label{fig1}
\end{figure*}

The relevant Feynman diagrams are given in Fig.~\ref{fig1}. All four diagrams contribute to the matrix element $\mathcal{M}(\phi\phi\rightarrow h_{\mu\nu}h_{\mu\nu})$   at order $\mathcal{O}(\kappa^2)$. The Feynman rules and details about each amplitude will be provided in the Appendix~\ref{app:feynrules}. Here we note that the graviton polarization vector is simply the product of two photon polarization vectors:
\begin{equation}
    \varepsilon^{\pm 2}_{\mu\nu}=\varepsilon_\mu^\pm\varepsilon_\nu^\pm\,,
\end{equation}which is transverse, traceless, symmetric in $(\mu\nu)$ and forms an orthonormal basis. For the squared amplitude, we will need the polarization sum:
\begin{equation}
    \sum_{\text{pol}}\varepsilon^{*\mu\nu} \varepsilon^{\alpha\beta}=\frac{\hat{\eta}^{\mu\alpha}\hat{\eta}^{\nu\beta}+\hat{\eta}^{\mu\beta}\hat{\eta}^{\nu\alpha}-\hat{\eta}^{\mu\nu}\hat{\eta}^{\alpha\beta}}{2} \,,
\end{equation}
with
\begin{equation}
    \hat{\eta}^{\mu\nu}\equiv \eta^{\mu\nu}-\frac{\upsilon^\mu\Bar{\upsilon}^\nu+\upsilon^\nu\Bar{\upsilon}^\mu}{2E_\omega^2}\,,
\end{equation}
and $\upsilon=(E_\omega,\Vec{\upsilon})$, $\Bar{\upsilon}=(E_\omega,-\Vec{\upsilon})$ for a graviton of energy $E_\omega$ and momentum $\Vec{\upsilon}$. The inflaton condensate is by construction at rest, and for a quadratic potential, its  four momentum is $p_1^\mu =p_2^\mu = (m_\phi,\Vec{0})^\mu$ where $m_\phi$ is the inflaton mass. Momentum conservation then implies $k_1^\mu=(m_\phi,\Vec{\upsilon})^\mu$, $k_2^\mu=(m_\phi,-\Vec{\upsilon})^\mu$ with $\Vec{\upsilon}\cdot \Vec{\upsilon}=m_\phi^2$.  In this setup, the diagrams (a) (b) in Fig.~\ref{fig1} vanish,\footnote{Note that in \cite{Ghiglieri:2022rfp,Ghiglieri:2024ghm}, the production of  gravitational waves from the thermal plasma is considered, which involves similar diagrams as shown in Fig.~\ref{fig1}, but there, the diagrams (a) (b) do contribute because the initial states are not at rest.} and the remaining two add up to yield the following squared amplitude summed over polarizations:
\begin{equation}
     \sum_{\text{pol}}|\mathcal{M}|^2=\frac{2m_\phi^4}{M_P^4}\times\frac{1}{4}\label{amplitudek2}\,.
\end{equation}
The factor $1/4$ accounts for the symmetry factors of the initial and final states. The rate of production of gravitational waves is 
\begin{equation}
    \Gamma_{ h} =\frac{\rho_\phi}{m_\phi}\frac{\sum_{\text{pol}}|\mathcal{M}|^2}{32\pi m_\phi^2}=\frac{2\rho_\phi m_\phi}{64\pi M_P^4}\label{ratek2}\,,
\end{equation}
where $\rho_\phi$ is the energy density of the inflaton, and the factor of 2 accounts for the production of 2 gravitons per interaction.

\section{Gravitational Wave Spectrum}
\label{spect}

The density of gravitational waves, $\rho_{\rm GW}$, produced from inflaton scattering is determined from the Boltzmann equation 
\beq
\dot{\rho}_{\rm GW} + 4H \rho_{\rm GW} =  (1+w_{\phi})\Gamma_{h}(t)\rho_{\phi}\, ,
\label{BGW}
\eeq
where $H$ is the Hubble parameter, $\Gamma_h$ is the rate for  producing gravitational waves computed in the previous section, and the equation of state parameter is
\beq\label{eq:wk}
w_{\phi} \;=\; \frac{k-2}{k+2}\,.
\eeq
 The Boltzmann equation (\ref{BGW}) is coupled to the corresponding equation for the evolution of $\rho_\phi$,
\beq
\dot{\rho}_{\phi} + 3H(1+w_{\phi})\rho_{\phi} \;\simeq\; -\Gamma_{\phi}(1+w_{\phi})\rho_{\phi} \, ,
\eeq
where $\Gamma_\phi$ is the inflaton decay rate which is responsible for reheating. For $\Gamma_\phi \ll H$ and
using $\frac{d}{dt} = a H \frac{d}{da}$,
where $a$ is the cosmological scale factor, this equation can be easily integrated to give
\begin{eqnarray}
\rho_\phi(a) & = & \rho_{\rm end} \left(\frac{a_{\rm end}}{a} \right)^{\frac{6k}{k+2}} \label{rhophik} \\
& = & \rho_{\rm end} \left(\frac{a_{\rm end}}{a} \right)^{3} \qquad  k=2\,. 
\label{Eq:rhophi}
\end{eqnarray}
Here $a_{\rm end}$ is the value of the scale factor when inflation ends, i.e., when $\ddot a = 0$ and $\rho_{\rm end}$ is the inflaton energy density at $a = a_{\rm end}$. This evolution is valid until $H \simeq \Gamma_\phi$.
At this time, decays become rapid and the density of inflatons becomes exponentially suppressed. 
In this section, we will fix $k=2$ corresponding to $w_\phi=0$.

Finally, we need to determine the reheating temperature which we assume is a result of inflaton decay. The density of the radiation bath is also given by a Boltzmann equation
\beq
\dot{\rho}_{R} + 4H \rho_{R} \;\simeq\; (1+w_{\phi})\Gamma_{\phi}(t)\rho_{\phi}\, ,
\eeq
For $k=2$, when $a \gg \aend$,
the solution to this is \cite{GKMO2}
\beq
\rho_R = \frac25 \frac{\Gamma_\phi \rhoend}{H_{\rm end}} \left( \frac{\aend}{a} \right)^\frac32 \, ,
\eeq
where $H_{\rm end} = \rhoend^\frac12/\sqrt{3} M_P$. 
Defining the ``moment'' of reheating when $\rho_\phi (\arh) = \rho_R(\arh)$
gives 
\beq
\trh = \alpha^{-\frac14} \left( \frac{2\sqrt{3}}{5} \Gamma_\phi M_P \right)^\frac12 \, ,
\label{trh}
\eeq
where $\alpha = g_{\rm RH} \pi^2/30$ and $g_{\rm RH}$ is the number of relativistic degrees of freedom at reheating. 
Here, we will not concern ourselves with the specifics of the reheating mechanism, but assume there is an appropriate coupling of the inflaton to matter leading to a decay rate, $\Gamma_\phi$, and the reheat temperature given in Eq.~(\ref{trh}).

The Boltzmann equation (\ref{BGW}) for the gravitational radiation can also be re-expressed as
\beq
\frac{1}{a^4} \frac{d (a^4 \rho_{\rm GW})}{da} = \frac{\Gamma_h} {a H}\rho_\phi \, ,
\eeq
which is easily solved using Eq.~(\ref{ratek2}) for $\Gamma_h$.
Remembering that $\Gamma_h \propto \rho_\phi$, we find that by integrating the Boltzmann equation from $a = \aend$ to $a = \arh$, the total energy density in gravitational waves is
\beq
\rho_{\rm GW}(\arh) = \frac{\sqrt{3} m_\phi \rho_{\rm end}^\frac32}{16 \pi M_P^3} \left( \frac{\aend}{\arh} \right)^4 \, ,
\eeq
or using $\rhorh = \alpha \trh^4 = \rhoend (\aend/\arh)^3$, we have
\beq
\rho_{\rm GW}(\arh) = \frac{\sqrt{3} \alpha^\frac43 m_\phi \rho_{\rm end}^\frac16 \trh^\frac{16}{3}}{16 \pi M_P^3} \, .
\label{rgwtot}
\eeq

In order to compare this result with the current and future experimental limits, we are interested in the frequency distribution of this gravitational wave background.
Because the source for the Boltzmann equation for $\rho_{\rm GW}$ in Eq.~(\ref{BGW}) is proportional to $a^3 \rho_\phi^2/H \propto a^{-\frac32}$, the bulk of the energy density in the background was produced immediately after inflation ends, at $a=\aend$.
These gravitons are produced monochromatically with energy $E_h=m_\phi$.  The frequency of these waves at $a = \arh$ is redshifted by $(\aend/\arh)$ and then further redshifted to today by another factor of $(\arh/a_0) =\xi\,(T_0/\trh)$, where $T_0$ is the temperature of the cosmic microwave background today. The factor $\xi$ is due to entropy conservation and  $\xi\equiv(g_0/g_{\rm RH})^{1/3}$
with $g_{\rm RH} = 427/4$ and $g_0 = (43/4)(4/11)$ 
so that $\xi \simeq 0.332$. 
Thus most of the gravitational waves are produced with 
frequency $2 \pi f_e = m_\phi (\aend/a_0)$, where the subscript on $f$ denotes those waves produced at $a=\aend$.

Despite the drop in the inflaton energy density as the Universe expands, gravitational waves continue to be produced (initially at a frequency $2 \pi f = m_\phi$) until
$H \simeq \Gamma_\phi$ when the inflaton density begins to decrease exponentially and the production of new gravitational waves ceases. 
The frequency today of a gravitational wave produced at some value of the scale factor, $a$, is then $2 \pi f = m_\phi (a/a_0)$ and the fraction of gravitational waves with frequency between $f$ and $f + df$ evaluated at $a = \arh$ is
\beq
\frac{d\rho_{\rm GW}}{df} =  \frac{\sqrt{3} \alpha^\frac32  \trh^6}{16 M_P^3} \left(\frac{m_\phi}{2 \pi f} \right)^\frac32   \left(\frac{T_0}{\trh} \right)^\frac12  \sqrt{\xi} \,  .
\eeq
Integrating this expression over $f$
between $f_e = (m_\phi/2\pi) (\aend/a_0)$ and $f_{\rm RH} =  (m_\phi/2\pi) (\arh/a_0) $ results in Eq.~(\ref{rgwtot}). The integration limits result from recognizing that no frequencies below $f_e$ are ever produced and the density with frequencies above $f_{\rm RH}$ are cut off exponentially as the inflaton decays. More precisely we have,
\begin{eqnarray}
    f_e & = & \frac{m_\phi}{2 \pi}\, \alpha^\frac13 \,\trh^\frac13 \,\xi \,T_0 \,\rhoend^{-\frac13} \simeq 4 \times 10^6~{\rm Hz}  \nonumber \\ 
    & \times & \left(\frac{m_\phi}{3\times 10^{13}~{\rm GeV}}\right) \left( \frac{\trh}{10^{10}~{\rm GeV}} \right)^\frac13 \nonumber \\ &\times& \left(\frac{(5.5 \times 10^{15}~{\rm GeV})^4}{\rhoend} \right)^\frac13\,.
    \label{fe}
\end{eqnarray}

 We are now in a position to compute the relative contribution to $\Omega_{\rm GW} h^2=(d\rho_{\rm GW}/d\ln f)/(\rho_{c,0}h^{-2})$, 
 \begin{eqnarray}
\Omega_{\rm GW} h^2  & = & \frac{h^2}{\rho_c} \frac{\sqrt{3} \alpha^\frac43 \trh^\frac{16}{3} m_\phi \rhoend^\frac16}{32 \pi M_P^3 } \left( \frac{f_e}{f} \right)^\frac12 
\xi^{4}\,\left( \frac{T_0}{\trh} \right)^4 \cr\cr
& \times&  \Theta(f-f_e) e^{-(5\sqrt{\alpha}/4) f^2/f_{\rm RH}^2}\cr\cr
&= & 1.3\times10^{-24}\left( \frac{f_e}{f} \right)^\frac12 \left(\frac{T_\text{RH}}{10^{10}\text{ GeV}}\right) ^\frac{4}{3}\cr\cr
&\times& \left(\frac{\rhoend} {(5.5 \times 10^{15}~{\rm GeV})^4}\right)^\frac16  \left(\frac{m_\phi}{3\times 10^{13}~{\rm GeV}}\right)\cr\cr
&\times&  \Theta(f-f_e) e^{-(5\sqrt{\alpha}/4) f^2/f_{\rm RH}^2}\,,
\label{ogw}
\end{eqnarray}
where the factor $\xi^{4} (T_0/\trh)^4$ accounts for the redshifting of the spectrum from $a=\arh$ to today and $\rho_c \simeq 8.1\times 10^{-47} h^2$~GeV$^4$; $h = H_0/100$~km/Mpc/s is the scaled present-day Hubble parameter.
We have also included the step function indicating the minimal frequency is $f_e$ and the approximate exponential cutoff at frequencies larger than $f_{\rm RH} = (\rhoend/\rhorh)^\frac13 f_e = (m_\phi/2 \pi) \xi (T_0/\trh) \simeq 5.6 \times 10^{13}$~Hz for the same set of normalizations. 

We can now apply this result to the inflationary models we described by the potentials in Eqs.~(\ref{staropot}) and (\ref{eq:Vattractor}).
Starting with the Starobinsky potential, 
we recall that the inflaton mass is determined by the normalization of the CMB anisotropy spectrum \cite{eno6},
\beq
m_\phi^2 \simeq \frac{24 \pi^2 A_s}{N_*^2}M_P^2 \, ,
\eeq
where $A_s = 2.1\times10^{-9}$ from the quadrupole normalization \cite{Planck}, and $N_*$ is the number of e-foldings from the Planck pivot scale ($k_* = 0.05$~ Mpc$^{-1}$) to the end of inflation. For $N_* = 55$, this gives $m_\phi \simeq 3 \times 10^{13}$ GeV. We also need to determine $\rhoend$, which follows from the determination of $\phi_{\rm end}$.  Recall that $\phi_{\rm end}$ is defined as the value of the inflaton background field value when
the Universe exits the phase of exponential expansion, when $\ddot{a}= 0$. For the Starobinsky potential, $\phi_{\rm end}$ is given by \cite{egno5,egnov}
\beq
\phi_{\rm end} = \sqrt{\frac{3}{2}} \ln\left[\frac{2}{11} \left(4 + 3\sqrt{3} \right)\right] M_P \simeq 0.63 M_P \, .
\eeq

 \begin{figure}[!t] 
    \centering
\includegraphics[width=0.5\textwidth]{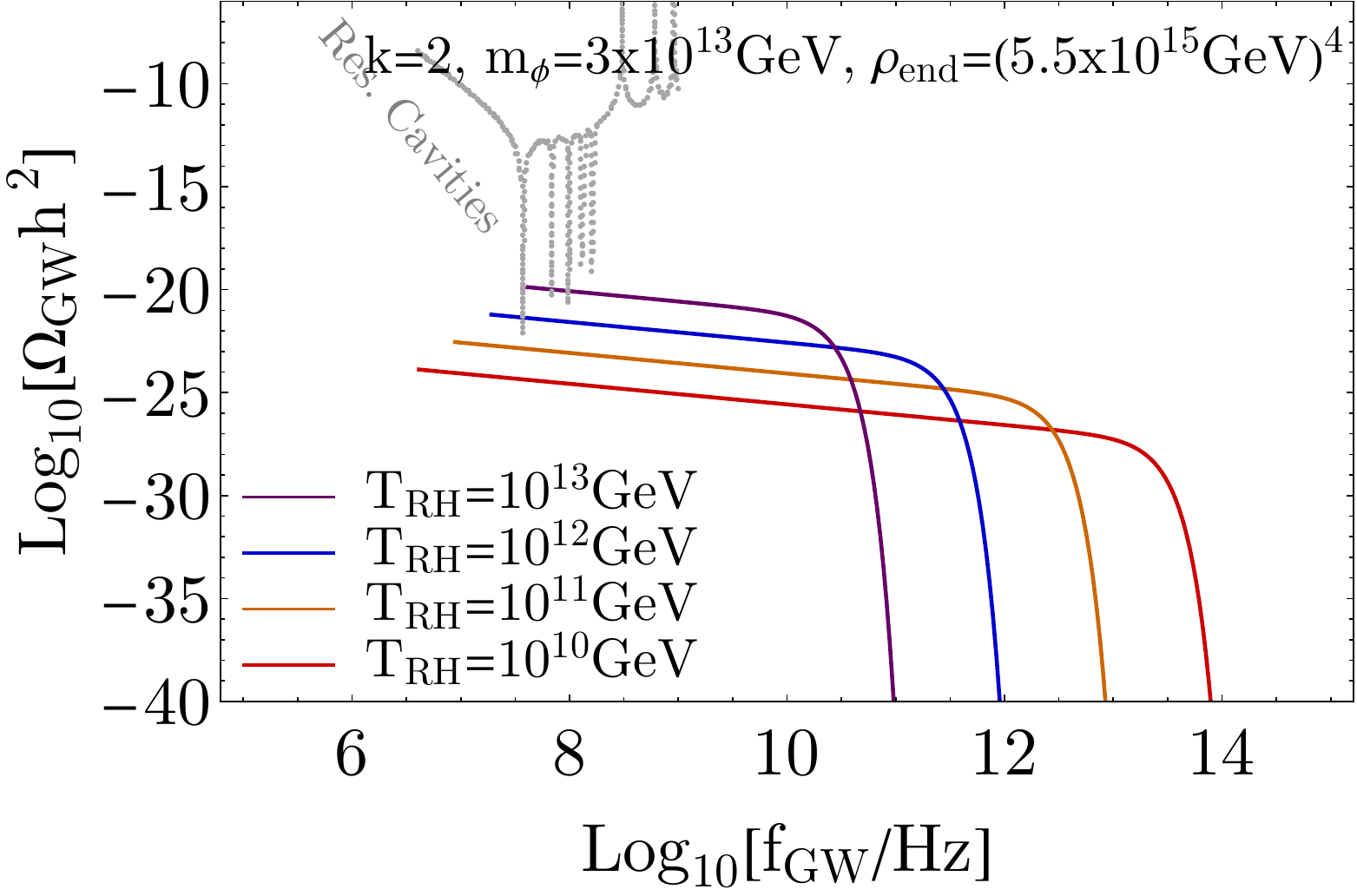}
  \caption{Gravitational wave spectra in Eq.~(\ref{ogw}) resulting from the inflaton condensate pair annihilation in the Starobinksy model with $m_{\phi}=3\times10^{13}{\rm GeV}$ and $\rho_{\rm end}=(5.5\times10^{15}{\rm GeV})^{4}$. Each solid line of different colors is differentiated by the specified distinct reheating temperature. The gray dotted line shows the sensitivity of the resonant cavities proposal~\cite{Herman:2020wao,Herman:2022fau}.}
    \vspace*{-1.5mm}
    \label{fig2}
    \end{figure}

When inflation ends, $\dot{\phi}_{\rm end}^2 = V(\phi_{\rm end})$ and $\rhoend = \frac32 V(\phi_{\rm end}) \simeq 0.175 m_\phi^2 M_P^2 \simeq (5.5 \times 10^{15}~{\rm GeV} )^4$. We have used these normalizations to determine $f_e$ in Eq.~(\ref{fe}). 

The prompt gravitational wave spectrum from Eq.~(\ref{ogw}) using the Starobinsky model inputs is shown in Fig.~\ref{fig2}. Shown is the contribution to $\Omega_{\rm GW} h^2$ for four different assumed reheat temperatures, $\trh = 10^{10}, 10^{11}, 10^{12}$, and $10^{13}$~GeV. For each temperature, the spectrum begins at a different frequency as $f_e$ has a slight dependence on $\trh$ as seen in Eq.~(\ref{fe}). At higher frequencies, we see the cut-off in the spectrum due to inflaton decay. Note that the energy density of the inflaton condensate becomes exponentially suppressed near reheating, and the production of gravitons from inflaton scattering effectively ceases. This is encoded in the exponential factor in Eq.~(\ref{ogw}), which makes the cutoff of $\Omega_{\rm GW}h^{2}$ sensitive to the choice of $T_{\rm RH}$. Therefore, $T_{\rm RH}$ can be independently read-off from $f_{e}$ and $f_{\rm RH}$ and this helps us extract values of $(m_{\phi},\rho_{\rm end})$ from the strength of $\Omega_{\rm GW}h^{2}$ as $\Omega_{\rm GW}h^{2}$ depends on $m_{\phi}$, $\rho_{\rm end}$ and $T_{\rm RH}$. Also shown as the gray dotted line is the potential sensitivity of the proposed GW detector using resonant cavities~\cite{Herman:2020wao,Herman:2022fau}.

Similarly, for the T-models with $k=2$ we can approximate the coupling $\lambda$ \cite{egnov}
\beq
\lambda \simeq \frac{3 \pi^2 A_s}{N_*^2}\,,
\eeq
and 
$\phi_{\rm end}$ is given by \cite{egnov}
\beq
\phi_{\rm end} =  \sqrt{\frac{3}{2}}  \ln\left[\frac{1}{11} \left(14 + 5\sqrt{3} \right)\right] M_P \simeq 0.88 M_P \,.
\eeq
In this case, the inflaton mass is slightly lower, $m_\phi \simeq 2.2 \times 10^{13}$~GeV and $\rhoend \simeq (5.2 \times 10^{15}~{\rm GeV} )^4$. This gives $f_e \simeq 3.2 \times 10^6$~Hz and $f_{\rm RH} \simeq 4.1 \times 10^{13}$~Hz. We find that $\Omega_{\rm GW}h^{2}$ from T-model inflaton scattering is indistinguishable (on the scale of the plot) from the Starobinsky model in Fig.~\ref{fig2} because of the similarity in $(m_{\phi},\rho_{\rm end})$ between the two models.

\section{Generalized Potentials}
\label{kne2}
In the previous section, we computed the gravitational wave spectrum from an inflaton potential with a quadratic minimum. More generally, the T-models admit potentials which can be expanded about the minimum with the form in Eq.~(\ref{kexp}). The gravitational wave spectrum will then also depend on the power $k$ and we analyze its effect in this section. 

The time dependence of the inflaton energy density for $a > \aend$ is given by the solution of Eq.~(\ref{rhophik}). We can parametrize the time dependence of the inflaton by
\beq
\phi(t) = \phi_0(t) \mathcal{P}(t)\,,
\eeq
where the function $\mathcal{P}(t)$ is quasi-periodic and characterizes the (an)harmonicity of the short time-scale oscillations in the potential. The amplitude $\phi_0(t)$ is characterized by the solution Eq.~(\ref{rhophik}) and varies on longer time-scales.

The inflaton oscillations can be understood by writing $V(\phi)=V(\phi_0)\cdot\mathcal{P}(t)^k$, where $\mathcal{P}(t)$ is expanded as a Fourier series:
\begin{equation}
  \mathcal{P}(t)=\sum_{n=-\infty}^{\infty}\mathcal{P}_n e^{-in\omega t}\,.
\end{equation}
Solving the equation of motion leads to \cite{GKMO2}
\begin{equation}
    \omega=m_\phi \sqrt{\frac{\pi k}{2(k-1)}}\frac{\Gamma\left(\frac{1}{2}+\frac{1}{k}\right)}{\Gamma\left(\frac{1}{k}\right)}\,.
\end{equation}
The inflaton mass and energy density are 
\begin{eqnarray}
m_\phi^2=V^{\prime\prime }(\phi_0),\quad \rho_\phi=V(\phi_0)=\frac{1}{2}\Tilde{m}_\phi^2\phi_0^2\label{rhomk}\,.
\end{eqnarray}
where we defined\begin{equation}
    \Tilde{m}_\phi^2\equiv \frac{2m_\phi^2}{k(k-1)}.
\end{equation} 
The Feynman rules for computing the diagrams in Fig.~\ref{fig1} are the same, except that the energy of the $n$-th oscillation mode is  $E_n=n\omega$ (and $\omega =  m_\phi$ for $k=2$), and the total four-momentum  of the condensate is $p_n=p_1+p_2=\sqrt{s}= (E_n,\Vec{0})$. For each oscillation mode $n$, we replace the inflaton legs $\phi^2$ by $\phi_0^2\left(\mathcal{P}^k\right)_n$, treating $\left(\mathcal{P}^k\right)_n$ as an interaction coefficient. Here, $\left(\mathcal{P}^k\right)_n$ is the Fourier coefficient of $\mathcal{P}(t)
^k$. The final total amplitude is obtained by summing each $|\mathcal{M}_n|^2$ over $n$.

We find that when summed over polarizations of outgoing gravitons, the squared amplitude for mode $n$ equals that of the $k=2$ case in Eq.~\eqref{amplitudek2} multiplied by an interaction coefficient:  
\begin{equation}
\sum_{\text{pol}}|\mathcal{M}_n|^2= \phi_0^4| (\mathcal{P}^k )_n|^2\frac{\Tilde{m}_\phi^4}{2M_P^4} \,.
\end{equation}

The energy transfer  rate per unit spacetime volume ($\text{Vol}_4$) is defined as\footnote{In practice, the sum starts at $n=2$ because two inflatons scatter, and only Fourier coefficients with even $n$ are non-vanishing.}
\begin{widetext} 
\begin{equation} \begin{aligned}\frac{\Delta E}{  \text{Vol}_4}\equiv (1+w_\phi)\Gamma_h\rho_\phi=\sum_{n=1}^\infty\int \frac{d^3 p_A}{(2\pi)^32p_A^0}\frac{d^3 p_B}{(2\pi)^32p_B^0}(p_A^0+p_B^0)\sum_{\text{pol}}|\mathcal{M}_n|^2(2\pi)^4\delta^4(p_{\phi,1}+p_{\phi,2} -p_A-p_B)\,.
\label{Etransf}
\end{aligned}
\end{equation}
\end{widetext}
We then obtain the energy transfer rate: 
\begin{equation}
    (1+w_\phi)\Gamma_h\rho_\phi= \frac{\rho_\phi ^2\omega}{4\pi M_P^4}\Sigma^k\,,\label{transrategenk}
\end{equation}
where we replaced $\phi_0$ using Eq.~\eqref{rhomk} and  defined
\beq
 \Sigma^k=\sum^\infty_{n=1} \Sigma^k_n \equiv\sum^\infty_{n=1}n| (\mathcal{P}^k )_n|^2 \, .
 \eeq
When  $k=2$, the solution for $\phi$ is approximately $\phi(t)=\phi_0 \cos(\omega t)$ where $\omega=m_\phi $, and the energy density is $\rho_\phi=m_\phi^2 \phi_0^2/2$. Then for positive energy, the non-vanishing Fourier coefficient  is $(\mathcal{P}^2)_{2}=1/4$ and thus $\Sigma^2=1/8$. Consequently, for $k=2$, we correctly reproduce the rate in Eq.~\eqref{ratek2}.  For $k=4~(6)$, $\Sigma^k$ is approximately $0.141~(0.146)$. In a complete analysis, one should take into account the contribution of each  mode $n$ to the power spectrum, which is weighted by the coefficient $\Sigma^k_n$. Note that the graviton carries the energy $n\omega/2$, so the sum over all modes gives rise to very high frequency waves in addition to the dominant $n=2$ mode. The Fourier coefficients decrease rapidly with $n$ if $k$ is not too large ($k\lesssim 10$). In practice, as we will see, the sum is mostly dominated by the first mode ($n=2$) and the subleading modes ($n>2$) affect a tail of the spectrum at higher frequencies that gradually falls off in the frequency.

For the T-models with $k>2$, an approximate solution for $\phi_{\rm end}$ is \cite{GKMO2}
\beq
\phi_{\rm end} \;=\;\sqrt{\frac{3}{8}}\, M_P \ln\left[ \frac{1}{2} + \frac{k}{3}\left(k+\sqrt{k^2+3}\right) \right]\,.
\eeq
The CMB normalization is generalized to 
\beq
\lambda = \frac{18 \pi^2 A_s}{6^\frac{k}{2} N_*^2}\,,
\eeq
giving $\lambda = 3.4 \times 10^{-12}~(5.7 \times 10^{-13}$) for $k=4~(6)$. These values then determine $\rhoend^\frac14 = 4.8 \times 10^{15}~{\rm GeV}~(4.8 \times 10^{15}~{\rm GeV})$.

The frequency spectrum is obtained in the same way as in the previous section.  At the end of reheating, the graviton energy density is:\begin{equation}
 \begin{aligned}
     \rho_\text{GW}(a_\text{RH})=&\frac{\sqrt{3} (k+2)\gamma_k}{8(4k-7)\pi} \Sigma^k \\&\times\rho_\text{end}^\frac{4k-7}{3k}M_P^\frac{4-4k}{k}(\alpha T_\text{RH}^4)^\frac{4+2k}{3k}\,,
 \end{aligned}   \label{rhoRHgenk}
\end{equation} where 
\begin{equation}
    \gamma_k \equiv \sqrt{\frac{\pi}{2}}k\frac{\Gamma\left(\frac{1}{2}+\frac{1}{k}\right)}{\Gamma\left( \frac{1}{k}\right)} \lambda^\frac{1}{k}\,.
\end{equation}
Note that in Eq.~(\ref{rhoRHgenk}),
the contribution of all modes (all values of $n$) have been included. 
As we mentioned before, for general $k$, the energy of the gravitational waves at the moment of production is not always  $m_\phi$, but given by  $E_\omega\simeq\omega$ of the dominant $n=2$ mode. In this case, the frequency today of the gravitational wave produced at $a$ is:
\begin{equation}
    f(a)=\frac{\omega}{2\pi} \frac{a}{a_0}=\frac{\gamma_k M_P^{\frac{4-k}{k}}\rho_\text{end}^\frac{k-2}{2k}}{2\pi a_0}a_\text{end}^\frac{3k-6}{k+2}a^\frac{8-2k}{k+2}\label{fagenk}\,.
\end{equation}
For $k=2$, we recover the relation $f\propto a$, while for $k=4$, $f$ is constant in $a$, meaning that the gravitational wave spectrum is almost monochromatic today for each harmonic mode, up to oscillation effects that we have neglected so far. If we sum up all inflaton modes, the $k=4$ spectrum will then feature multiple peaks with increasing frequency and decreasing intensity.  Finally, when $k>4$, $f$ decreases with $a$, hence the observed  frequency of gravitational waves produced at the end of inflation is higher than that produced at the end of reheating.

Let us start with $k>4$. Requiring the integral of $d\rho_{\rm GW}/df$ from $f_\text{RH}$ to $f_e$ to coincide with Eq.~\eqref{rhoRHgenk} gives for the dominant $n=2$ mode: \begin{equation} \begin{aligned}\frac{d\rho_\text{GW}}{df}&= (2\pi f)^\frac{3k-3}{k-4}T_\text{RH}^\frac{3+4k}{k}(\xi T_0)^\frac{4k-7}{4-k}
\\& \times \frac{\sqrt{3} (k+2)}{4 (k-4)M_P^\frac{3}{k}} \gamma_k^{\frac{3-3k}{k-4}} \alpha^\frac{3(k+2)}{2k(4-k)}\Sigma^k_2 \,.\end{aligned}   \end{equation}

Accordingly, we find the resulting gravitational wave spectrum for $n=2$\footnote{For $n=2$ and $k=6$, the contribution is $\Sigma^6_2 = 0.083$ compared with $\Sigma^6 = 0.146$. }: 
\begin{equation}
\begin{aligned}\Omega_\text{GW}h^2&=\frac{h^2}{\rho_c} \left(\frac{f}{f_e}\right)^{\frac{4k-7}{k-4}}\gamma_k\Sigma^k_2\\& \times \left(\frac{\xi T_0}{T_\text{RH}}\right)^4 T_\text{RH}^\frac{16+8k}{3k} \rho_\text{end}^\frac{4k-7}{3k} \alpha^\frac{2k+4}{3k}\\& \times\frac{\sqrt{3}M_P^{\frac{4-4k}{k}}}{8\pi(k-4)} (k+2)\times\Theta(f_e-f)\times C(f)\\  = & 1.4\times10^{-15}\left( \frac{f}{f_e} \right)^{\frac{17}{2}} \left(\frac{T_\text{RH}}{10^{10}\text{ GeV}}\right) ^{-\frac{4}{9}} \\&\times\left(\frac{\rhoend} {(4.8 \times 10^{15}~{\rm GeV})^4}\right)^\frac{17}{18}\\&\times  \Theta(f_e-f)\times C(f)\quad(k=6)\,.
\label{ogw2}
\end{aligned}
\end{equation}
where
\begin{equation}
C(f)=
\Theta(f-f_{\rm RH})+\Theta(f_{\rm RH}-f)\left(\frac{f}{f_{\rm RH}}\right)^{\frac{8k}{k-2}}\,.
\label{CF}
\end{equation}

The cut-off function $C(f)$ encodes how rapid $\rho_{\phi}(t)^{2}$ decreases near the reheating time.\footnote{For $k>2$, the cut off is no longer exponential, but falls over slower (as a power-law) for decays to fermions and faster for decays to scalars. This can be understood since the decay rate to fermions is proportional to the decreasing inflaton mass, and decays to scalars is inversely proportional to the decreasing inflaton mass.} Here we focus on the inflaton perturbative decay to a pair of fermions . For the decay to a pair of bosons, we find the cut-off to be faster than $C(f)$ since the decay rate increases in time. Note that the relation $\rho_{\phi}\propto (T/T_{\rm RH})^{4k/(k-2)}$ does not depend on a value of $k>4$ although a proportionality constant does~\cite{GKMO2}.\footnote{Given $\rho_{\phi}\propto t^{2k/(2-k)}$, $a\propto\sqrt{t}$ and $m_{\phi}\propto\rho_{\phi}^{(k-2)/2k}$ near the reheating~\cite{GKMO2}, one finds $\omega\propto m_{\phi}\propto a^{-2}$ and thus $T\propto a^{-1}\propto f$ which converts the temperature dependence of $\rho_{\phi}$ into the frequency dependence.}

\begin{figure}[t]
\centering
\hspace*{-5mm}
\includegraphics[width=0.49\textwidth]{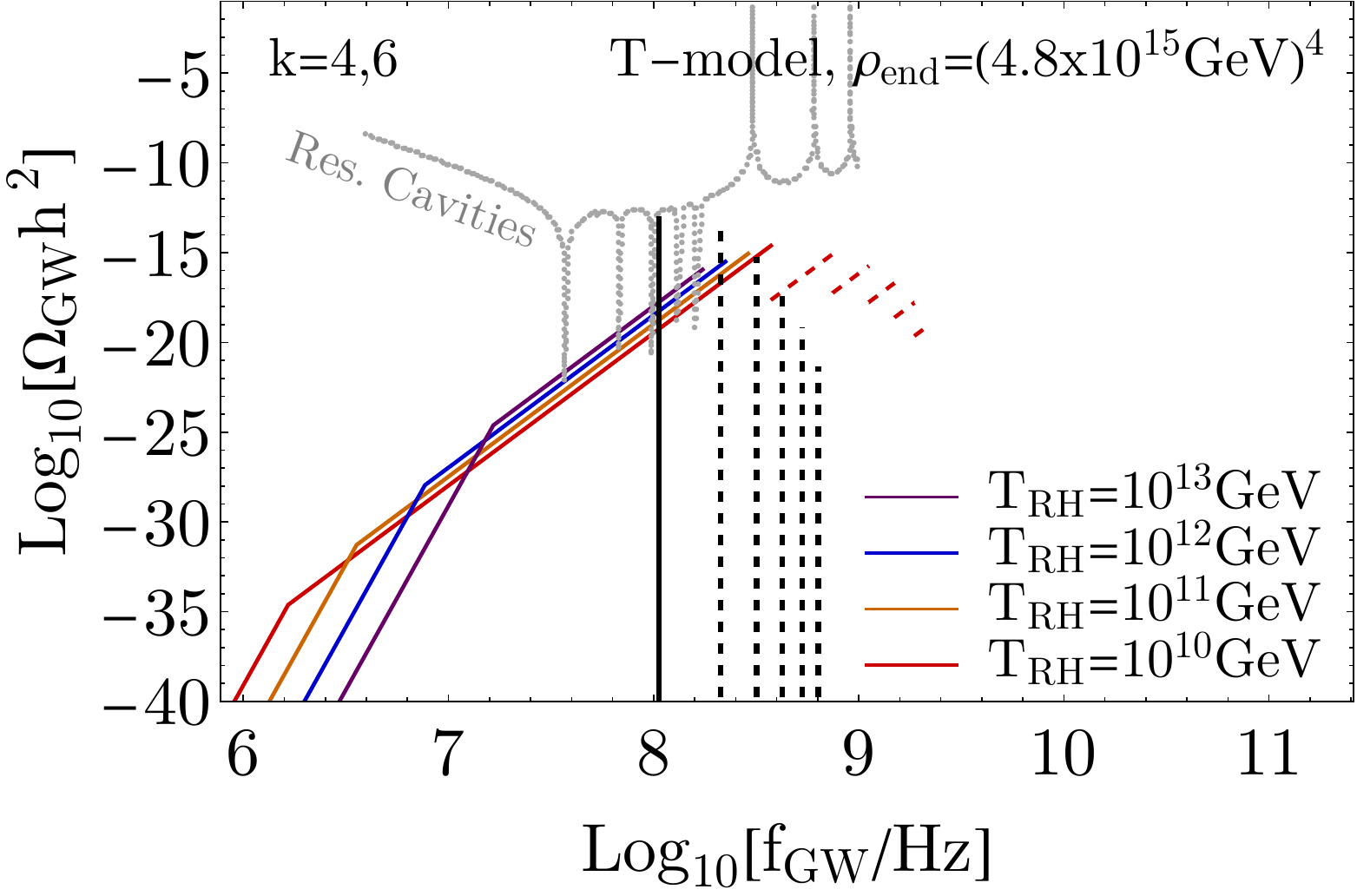}
\caption{Gravitational wave spectra in Eqs.~(\ref{ogw2}) (\ref{CF}) and (\ref{ogw3}) resulting from the inflaton condensate pair annihilation in the T-model   with $\lambda = 3.4 \times 10^{-12}~(5.7 \times 10^{-13}$) for $k=4~(6)$ and $\rho_{\rm end}=(4.8\times10^{15}{\rm GeV})^{4}$. Each solid line of different colors is differentiated by the specified distinct reheating temperature and shows the GW spectrum for 
$k=6$. The $k=4$ case with the narrow frequency range in Eq.~(\ref{fwidth}) is shown as the vertical black line. Solid lines show the contribution from $n=2$, and the dashed lines show the contributions for the higher modes $n=4,6,8,10,12$. The gray dotted line shows the sensitivity of the resonant cavities proposal~\cite{Herman:2020wao,Herman:2022fau}.}
\vspace*{-1.5mm}
\label{fig3}
\end{figure}

For general $k>4$, the spectrum increases with frequency as $\Omega_\text{GW}h^2\sim f^\frac{4k-7}{k-4}$,   which gives  $f^{\frac{17}{2}}$, $f^{\frac{25}{4}}$   for $k=6,8$ respectively. The frequency spectrum for $n=2$ is shown as colored solid lines in Fig.~\ref{fig3} for $k=6$. Here we see the decay cut off at lower frequencies in contrast to the case of $k=2$ where the high frequency waves were cut off. Re-iterating, for $k=2$, independent of the time the wave is produced, the only mode produced is $n=2$ with frequency $\omega = m_\phi$, which gets redshifted to today. Waves produced earlier are redshifted more, and appear at lower frequency today.
In contrast, for $k>4$, 
while the $n=2$ mode is produced with frequency $\omega$, $\omega$ is no longer constant and decreases in time. Indeed this decrease more than compensates for the decrease in redshift, and waves produced later now have lower observed frequency. For the case $k=4$, the change in mass (frequency) exactly compensates for the redshift and all waves are produced with the same frequency today.

So far, we have made the approximation that the waves are produced with energy $\omega$ corresponding to the dominant inflaton mode $n=2$. To study the effect of $n>2$ modes, one uses the energy transfer rate \eqref{transrategenk}, replacing the sum by $\Sigma^k_n$. The frequency \eqref{fagenk} becomes $f_n(a)=n\omega a /(4\pi a_0)$, and the same computation can be carried out to obtain the final spectrum summed over $n$. In Fig.~\ref{fig3}, for the reheating temperature $T_{\rm RH}=10^{10}{\rm GeV}$ and $k=6$, we show the spectra induced by the scattering of the higher mode inflaton condensates with $n=4,6,8,10,12$ as the red dashed lines. In principle, there are contributions from much higher modes, i.e. $k>12$. Since it is difficult to distinguish the higher mode given the log scale in the plot, we display only up to $n=12$. For $k=6$, the similar dashed lines of other colors corresponding to other reheating temperatures are expected to be present although we intentionally avoided to show them for the clarity of the plot.

For the case of $k=4$, as we just noted, 
according to Eq.~\eqref{fagenk}, the frequency observed today is a constant $f_4=\gamma_4 \xi T_0\alpha^{1/4} /(2\pi)\simeq 1.06\times 10^8 \text{ Hz} $ for $n=2$, and is independent of the reheating temperature. In reality, however, the spectrum is not exactly a Dirac delta function, but small fluctuations of the frequency -- due to the oscillations of $\phi_0$ -- will result in a  finite peak. The same reasoning holds for higher inflaton modes $n>2$, and we expect a series of peaks in the spectrum with frequency $n f_4/2$, with $n=4,6,8\cdots$. Using the Boltzmann equation, the energy density of the graviton at the end of reheating is, for $k=4$:
\begin{equation}\begin{aligned}
    \rho_\text{GW}(a_\text{RH})=& \frac{3\sqrt{3}\rho_\text{end}^\frac{3}{4}\alpha T_\text{RH}^4}{36\pi M_P^3}\Sigma^4\gamma_4
    \label{k4tot}
    \end{aligned}\, .
\end{equation}

The $n=2$ mode included in Eq.~(\ref{k4tot}) can be obtained from the integral:
\begin{equation}
  \rho_\text{GW}(a_\text{RH})=   \int_{f_4-\delta f}^{f_4+\delta f} df \frac{1}{f} F(f)\,,
\end{equation}
and other modes can be obtained by changing the integration bounds to $n f_4/2 \pm \delta f$. Here $F(f)\equiv d\rho_\text{GW}/ d \ln f $. For simplicity, the integrand is approximated to be constant over the interval  $( f_4-\delta f , f_4+\delta f)$. Therefore, 
\begin{equation}
     \rho_\text{GW}(a_\text{RH})\simeq 2\frac{\delta f}{f_4} F\,.
\end{equation}

\begin{figure}[t]
\centering
\hspace*{-5mm}
\includegraphics[width=0.49\textwidth]{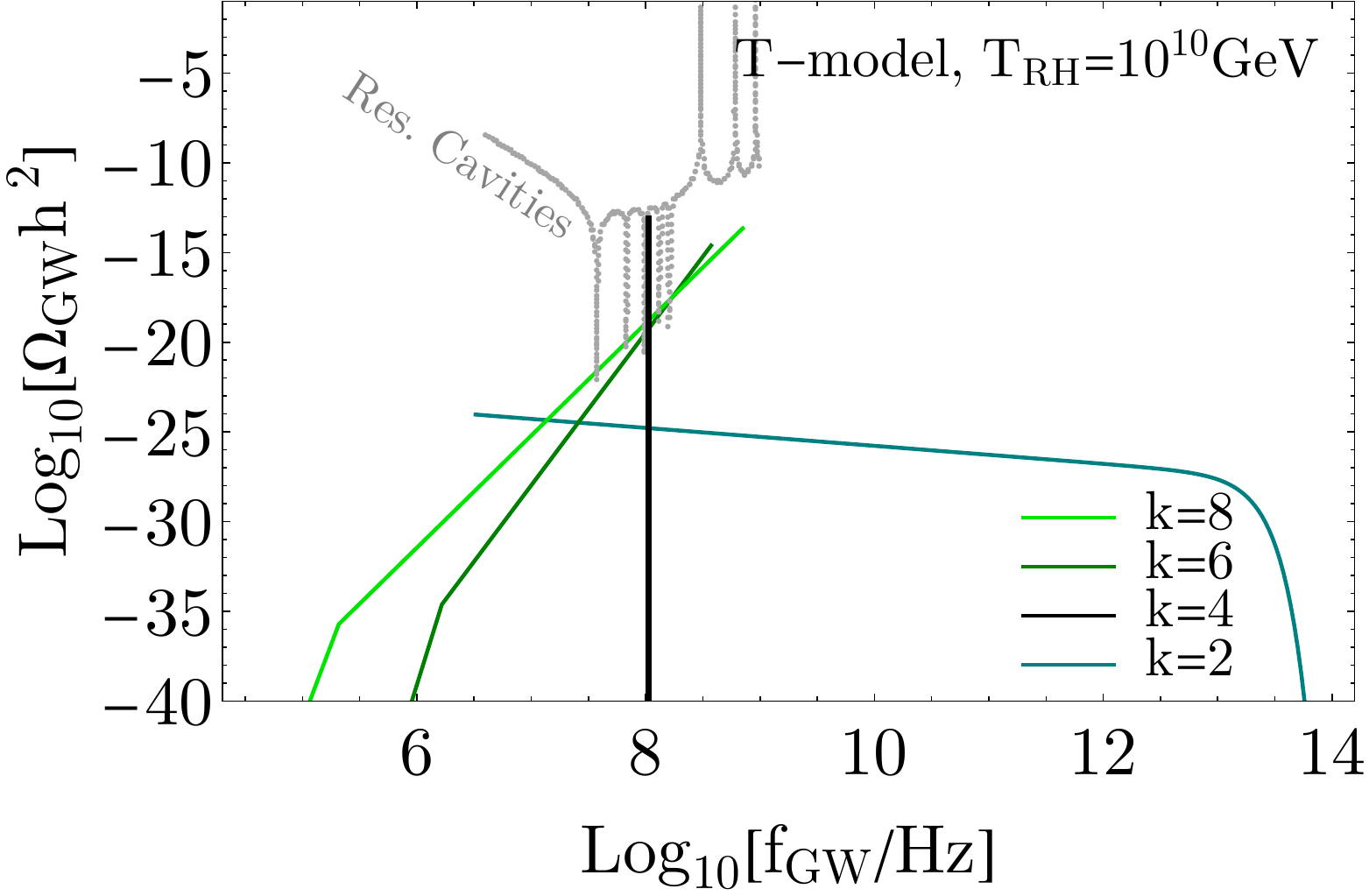}
\caption{Gravitational wave spectra from the dominant $n=2$ mode resulting from the inflaton condensate pair annihilation in various $k$'s in T-model (see Eq.~(\ref{kexp})) for a fixed $T_{\rm RH}=10^{10}{\rm GeV}$. The gray dotted line shows the sensitivity of the resonant cavities proposal~\cite{Herman:2020wao,Herman:2022fau}.}
\vspace*{-1.5mm}
\label{fig4}
\end{figure}

Typically, the fluctuation of $\phi_0$ scales as $\delta\phi_0 \sim H_\text{end}\sim 10^{-5} M_P$. As a result, the graviton energy oscillates as $\delta\omega=\gamma_4\lambda^\frac{1}{4} \delta\phi_{\rm end}  \simeq \gamma_4\lambda^\frac{1}{4}H _\text{end} $, which gives rise to the following frequency fluctuation:\begin{equation}
\frac{\delta f}{f_4}= \frac{\delta\omega}{\omega}=\frac{\delta{\phi_0}}{\phi_0}\simeq
10^{-5} \, .
\label{fwidth}
\end{equation} 
Then the height of the power spectrum for $n=2$ is evaluated from:
\begin{equation}
    \begin{aligned}
        \Omega_\text{GW}h^2 =&\frac{h^2}{\rho_c}\left(\frac{\xi T_0}{ T_\text{RH}}\right)^4 F \times\theta(f- f_4+\delta f)\theta(f_4+\delta f -f)\\=& \frac{h^2}{\rho_c}\frac{\left( {\xi T_0} \right)^4}{2(\delta f/ f_4)}\frac{3\sqrt{3}\rho_\text{end}^\frac{3}{4}\alpha  }{36\pi M_P^3}\Sigma^4_2\gamma_4\\&\times\theta(f- f_4+\delta f)\theta(f_4+\delta f -f)\\\simeq &7.7\times 10^{-14}\times\theta(f- f_4+\delta f)\theta(f_4+\delta f -f)\,,
    \end{aligned}
    \label{ogw3}
\end{equation}
which is independent of the reheating temperature. This (nearly) monochromatic spectrum is shown by the solid vertical line in Fig.~\ref{fig3}. Also shown are the higher mode contributions corresponding to $n=4,6,8,10,12$ as dashed vertical lines.

A comparison of the spectra for $k=2, 4, 6$, and 8 is shown in Fig.~\ref{fig4} for $n=2$ and $\trh = 10^{10}$~GeV. Here we see clearly the remarkable difference in the spectra determined by the inflaton potential.

\section{Summary}
\label{sec:sum}

One of the pressing problems confronting inflationary models is degeneracy among the few available observables. Most inflation models predict $\Omega = 1$ (i.e., they are nearly all degenerate in the curvature).
Among the T-models, all give nearly equivalent 
values of $n_s$ and $r$ for any value of $k$. The Starobinsky models give very similar results to the $k=2$ T-model. Breaking this degeneracy is of prime importance.

Here, we have computed the production of gravitational waves directly resulting from inflaton scattering within the condensate. This production is inevitable as it relies only on the interactions of the inflaton with gravity.  While the production mechanism is completely independent of any other interaction of the inflaton,
the final gravitational wave spectrum is redshifted from its initial production to the present day. This redshifting
depends on the epoch of reheating and hence on the interactions of inflation with the Standard Model.

However, we have shown that
unlike the predictions for the CMB anisotropy observables, the resulting gravitational wave spectrum is very sensitive to the model parameter, $k$. Indeed we have seen that for $k = 2$,
the intensity of the spectrum 
decreases from a maximum at low frequencies (those produced at the onset of reheating when the energy density is highest), to higher frequencies which are eventually cut off when the inflaton decays (and reheating is complete). This is shown in Fig.~\ref{fig2}.
In contrast, we have shown that for $k=4$, we expect a nearly monochromatic signal as
each wave produced between the end of inflation and reheating result in a common frequency ($\simeq 10^8$ Hz). In further contrast, for $k>4$, we have seen that $\Omega_{\rm GW}$ increases in frequency, with the maximum intensity now occurring at the largest frequencies (produced at the end of inflation) and the reheating cut off occurs at low frequencies. This is shown in Fig.~\ref{fig3}.
For fixed $T_{\rm RH}$, a comparison of the spectra for different $k$ is shown in Fig.~\ref{fig4}.

It is also important to note, that if in addition to a measurement of the shape of the spectrum (which could determine $k$), a measurement of the frequency at which the cut off  occurs can be directly related to the reheating temperature. This provides us with a rare potential signature of the reheating process.
Furthermore, if multiple harmonic modes are measured, these can be translated into a direct measurement of the inflaton mass!
To be fair,  the gravitational wave intensity predicted in these models is not large. While the prediction is robust, it is only 
within reach of future resonant cavity detectors. But we remain hopeful that improved technology will produce detectors that may probe this unique signal. 

We conclude by commenting on a similar spectrum for the cosmic axion background (axion dark radiation) resulting from the inflaton scattering. In the pre-inflationary scenario, the third vertex in Appendix~\ref{app:feynrules} applies to the axion as well. Then, there can be axion pair production from, for instance, the s-channel inflaton condensate scattering mediated by the graviton. Given that (1) the corresponding amplitude is of the same order of magnitude as that of graviton production and (2)  as long as the axion mass is smaller than $T_{0}$, the axion serves as dark radiation from the time of its production, and we expect the spectrum of axions $\Omega_{a} h^2=(d\rho_{a}/d\ln f)/(\rho_{c,0}h^{-2})$ from inflaton scattering to be very similar to $\Omega_{\rm GW} h^2$ which we have presented in this work. Although the future sensitivity of DMRadio in the frequency range of interest hardly reaches the expected $\Omega_{a} h^2$~\cite{Chaudhuri:2014dla,Silva-Feaver:2016qhh,Dror:2021nyr}, this new production mechanism may be of help in the future search for the cosmic axion background.

\vspace{0.5cm} 
\noindent
\acknowledgements
The authors thank Simon Cl\'ery, Marcos Garcia, and Yann Mambrini for valuable discussions during the completion of our work. This work was supported in part by DOE grant DE-SC0011842 at the University of Minnesota.

\appendix

\section{Feynman rules and amplitudes}
\label{app:feynrules}
We provide below the Feynman rules for the propagators and trilinear as well as quartic vertices \cite{Choi:1994ax}, generalized to arbitrary $k\geq2$. The curly lines represent the graviton and the dashed lines are the scalar.  
\begin{figure}[H]
\begin{tabular}{>{\centering\arraybackslash}m{1in}>{\centering\arraybackslash}m{1in}}
\centering
{\includegraphics[width=0.09\textwidth]{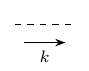}} & $$ =\frac{i}{k^2-\Tilde{m}_\phi^2}$$
\end{tabular}
\end{figure}
\begin{figure}[H]
\begin{tabular}{>{\centering\arraybackslash}m{1in}>{\centering\arraybackslash}m{1in}}
\centering
{\includegraphics[width=0.12\textwidth]{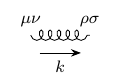}} & $$ \Pi^{\mu\nu\rho\sigma}=i\frac{\eta^{\mu\rho}\eta^{\nu\sigma}+\eta^{\mu\sigma}\eta^{\nu\rho}-\eta^{\mu\nu}\eta^{\rho\sigma}}{2k^2}$$
\end{tabular}\\\begin{tabular}{>{\centering\arraybackslash}m{1in}>{\centering\arraybackslash}m{1in}}
\centering
{\includegraphics[width=0.12\textwidth]{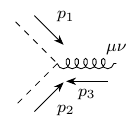}} & $$ \begin{aligned}t_{\phi,\mu\nu}=& \frac{i\kappa}{2}\left[p_{1\mu}p_{2\nu}+p_{1\nu}p_{2\mu} \right.\\&\left.-\eta_{\mu\nu}\left(p_1\cdot p_2 +\Tilde{m}_\phi^2\right)\right]\end{aligned} $$
\end{tabular}\\
\begin{tabular}{>{\centering\arraybackslash}m{1in}>{\centering\arraybackslash}m{1in}}
\centering{\includegraphics[width=0.11\textwidth]{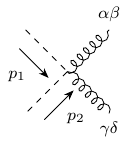}} & $$\begin{aligned}
    \tau_{\alpha\beta,\gamma\delta}=&-i\kappa^2 \left[I_{\alpha\beta,\rho\xi}I^{\xi}{}_{\sigma,\gamma\delta}\left(p_1^\rho p_2^\sigma +p_2^\rho p_1^\sigma \right)\right.\\&\left.-\frac{1}{2}\left(\eta_{\alpha\beta}I_{\rho\sigma,\gamma\delta}+\eta_{\gamma\delta}I_{\rho\sigma,\alpha\beta}\right)p_1^\sigma p_2^\rho\right.\\&\left.-\frac{1}{2}\left(I_{\alpha\beta,\gamma\delta} -\frac{1}{2}\eta_{\alpha\beta}\eta_{\gamma\delta}\right)\right.\\&\left.\times\left(p_1\cdot p_2 +\Tilde{m}_\phi^2\right)\right]
\end{aligned} $$
\end{tabular}\\\begin{tabular}{>{\centering\arraybackslash}m{1in}>{\centering\arraybackslash}m{1in}}
\centering
{\includegraphics[width=0.12\textwidth]{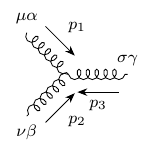}} 
& $$ \begin{aligned}V_{\mu\alpha,\nu\beta,\sigma\gamma}=&  i\kappa  \text{[Sym]}\left[  {P}_3\left(p_{1\sigma } p_{2\gamma} \eta_{\mu\nu}\eta_{\alpha\beta} \right)\right.\\&\left.-\frac{1}{2} {P}_6\left(p_1\cdot p_2 \eta_{\mu\alpha}\eta_{\nu\sigma}\eta_{\beta\gamma}\right)\right.\\&\left.+2 {P}_3\left(p_{1 }\cdot p_{2 } \eta_{\alpha\nu}\eta_{ \beta\sigma}\eta_{\gamma\mu} \right)\right.\\&\left.-\frac{1}{2}P_3\left(p_1\cdot p_2 \eta_{\mu\nu}\eta_{ \alpha\beta }\eta_{\sigma\gamma} \right) \right.\\&\left.-P_6\left(p_{1\sigma}p_{2\mu}\eta_{\alpha\nu}\eta_{\beta\gamma}\right)\right.\\&\left.+\frac{1}{4}P_3\left(p_1\cdot p_2 \eta_{\mu\alpha}\eta_{ \nu\beta }\eta_{\sigma\gamma} \right)\right]
\end{aligned} $$
\end{tabular}
\end{figure}
\noindent where $[\text{Sym]}$ means symmetrizing the indices $\mu\alpha,\nu\beta,\sigma\gamma$, and $P_i$ is the sum of all possible ($i$ terms) permutations among $(p_1\mu\alpha)$, $(p_2\nu\beta)$, $(p_3\sigma\gamma)$. We also defined 
\begin{equation}
    I_{\alpha\beta,\gamma\delta}=\frac{1}{2}\left(\eta_{\alpha\gamma}\eta_{\beta\delta}+\eta_{\alpha\delta}\eta_{\beta\gamma}\right) \, .
\end{equation}
Combining these Feynman rules, the first diagram in Fig.~\ref{fig1} evaluates to (for $k=2$):
\begin{equation}
\begin{aligned}
    \mathcal{M}_a=&-i\frac{\kappa^2}{4}\left[p_{1\mu}(k_{1\nu}-p_{1\nu})+p_{1\nu}(k_{1\mu}-p_{1\mu})\right.\\&\left.\quad -\eta_{\mu\nu}\left(p_1\cdot k_1-p_1\cdot p_1 +m_\phi^2\right)\right]\\&\times \left[p_{2\alpha}(p_{1\beta}-k_{1\beta})+p_{2\beta}(p_{1\alpha}-k_{1\alpha})\right.\\&\left.\quad -\eta_{\alpha\beta}\left(p_2\cdot p_1-p_2\cdot k_1 +m_\phi^2\right)\right]\\&\times \varepsilon^{\mu\nu}(k_1) \varepsilon^{\alpha\beta}(k_2)\times \frac{1}{(p_1-k_1)^2-m_\phi^2}\, .
\end{aligned}
\end{equation}
Meanwhile, for the inflaton condensate, there is no spatial momentum $p_1=(m_\phi,\Vec{0})$   hence the contractions $p_{1\mu}\varepsilon^{\mu\nu}$ or $p_{1\nu}\varepsilon^{\mu\nu}$ only pick up the zeroth components $\varepsilon^{0\nu}$ or $\varepsilon^{\mu0}$ which are 0 . In addition, the polarization vector is traceless $\eta_{\mu\nu}\varepsilon^{\mu\nu}=0$, therefore,
$\mathcal{M}_a=0$. The same reasoning holds for the second diagram, $\mathcal{M}_b=0$, and this is true also for general $k$.  The remaining two diagrams give the following non-vanishing contributions to the total squared amplitude, for $k=2$:
\begin{equation}
    \begin{aligned}
        &\sum_{\text{pol}}|\mathcal{M}_b|^2=\frac{9m_\phi^4}{2M_P^4}\quad  ,\quad \sum_{\text{pol}}|\mathcal{M}_c|^2=\frac{8 m_\phi^4}{M_P^4}  \, , \\&\sum_\text{pol}\left(\mathcal{M}_b\mathcal{M}_c^*+\mathcal{M}_b^*\mathcal{M}_c\right)=-\frac{12 m_\phi^4}{M_P^4}\, .
    \end{aligned} 
\end{equation}
The sum of them is then \eqref{amplitudek2}. For general $k$, each mode $n$ receives the contributions:\begin{equation}\begin{aligned}
\sum_\text{pol}    |\mathcal{M}_{c,n}|^2=&\phi_0^4 | (\mathcal{P}^k )_n|^2\frac{ (n^2\omega^2+2\Tilde{m}_\phi^2 )^2}{8M_P^4} \, ,
 \\
 \sum_\text{pol}   |\mathcal{M}_{d,n}|^2=&\phi_0^4 | (\mathcal{P}^k )_n|^2\frac{ (n^2\omega^2+4\Tilde{m}_\phi^2 )^2}{8M_P^4}\, ,\\\sum_\text{pol}
    \mathcal{M}_{c,n}^*\mathcal{M}_{d,n}+\text{c.c.}=&-\phi_0^4| (\mathcal{P}^k )_n|^2\\&\times \frac{ (n^2\omega^2+2\Tilde{m}_\phi^2 ) (n^2\omega^2+4\Tilde{m}_\phi^2 )}{4M_P^4}    \, .
\end{aligned}
\end{equation}


\end{document}